\documentclass[openacc]{rsproca_new}

\usepackage{lipsum}
\usepackage{multicol}
\usepackage{lmodern}
\bibliography{refs.bib}
\usepackage{float}
\usepackage{graphicx}

\usepackage{amsthm}

\renewcommand{\figurename}{Fig.}
\usepackage{amsmath}
\renewcommand{\thesubsubsection}{\arabic{section}.\arabic{subsection}.\alph{subsubsection}}
\makeatletter

\newcommand{\Rmnum}[1]{\expandafter\@slowromancap\romannumeral #1@}

\jname{rsif}
\Journal{J R Soc Interface\ }

\begin{document}

\title{Strategic competition in informal risk sharing mechanism  versus collective index insurance}

\author{
Lichen Wang$^{1}$, Shijia Hua$^{1}$, Yuyuan Liu$^{1}$, Zhengyuan Lu$^{1}$, Liang Zhang$^{1}$, Linjie Liu$^{1,2}$, Attila Szolnoki$^{3}$}

\address{$^{1}$College of Science, Northwest A\&F University, Yangling, Shaanxi, China\\
$^{2}$College of Economics \& Management, Northwest A \& F University, Yangling, Shaanxi 712100, China\\
$^{3}$Institute of Technical Physics and Materials Science, Centre for Energy Research, Budapest, Hungary}

\subject{evolution, computational mathematics}

\keywords{Evolutionary game theory, Markov process, Index insurance, Basis risk, Risk sharing pool}

\corres{
Linjie Liu\\
\email{linjie140@126.com}
}

\begin{abstract}

The frequent occurrence of natural disasters has posed significant challenges to society, necessitating the urgent development of effective risk management strategies. From the early informal community-based risk sharing mechanisms to modern formal index insurance products, risk management tools have continuously evolved. Although index insurance provides an effective risk transfer mechanism in theory, it still faces the problems of basis risk and pricing in practice. At the same time, in the presence of informal community risk sharing mechanisms, the competitiveness of index insurance deserves further investigation. Here we propose a three-strategy evolutionary game model, which simultaneously examines the competitive relationship between formal index insurance purchasing ($I$), informal risk sharing strategies ($S$), and complete non-insurance ($A$). Furthermore, we introduce a method for calculating insurance company profits to aid in the optimal pricing of index insurance products. We find that basis risk and risk loss ratio have significant impacts on insurance adoption rate. Under scenarios with low basis risk and high loss ratios, index insurance is more popular; meanwhile, when the loss ratio is moderate, an informal risk sharing strategy is the preferred option. Conversely, when the loss ratio is low, individuals tend to forego any insurance. Furthermore, accurately assessing the degree of risk aversion and determining the appropriate ratio of risk sharing are crucial for predicting the future market sales of index insurance.
\end{abstract}


\begin{fmtext}
\section{Introduction}
In the global economic environment, the frequent occurrence of natural disasters (such as floods and droughts) poses a severe threat to the economic stability and development of individuals, communities, and nations \cite{clarke2016dull,keim2020epidemiology,lesk2016influence,shaluf2007overview}. These disasters not only destroy infrastructure and undermine productivity but also lead to long-term social and economic instability, thereby exacerbating poverty and inequality \cite{adeagbo2016effects,cappelli2021trap,garcia2022disaster,hashim2016climate}. Therefore, constructing and implementing effective risk management strategies to mitigate the adverse effects of natural disasters has become a matter of urgency \cite{mora2009disasters,smolka2006natural}.

In the early stages of human society, informal risk sharing mechanisms provided farmers with a buffer against potential risks through interpersonal trust or contractual relationships \cite{HABTOM2007218,Mobarak2013}. For example, a community-based mutual aid cooperation mechanism, by establishing public funds pools, achieved centralized

\end{fmtext}


\maketitle

\begin{multicols}{2}
\noindent
management of resources, enabling individuals to receive timely support in emergencies, thereby effectively dispersing risks. This community-based mutual aid system not only strengthened the connections between members but also enhanced the ability to resist external shocks. These mechanisms rely on mutual assistance and resource sharing among community members. However, when faced with large-scale regional disasters such as floods or droughts, these mechanisms become inadequate as resource demands exceed their capacity, especially when most members of the group suffer significant losses at the same time. In addition, moral hazard, such as deliberately exaggerating losses in order to obtain more aid, is also an important obstacle that limits the widespread trust and continuous operation of these informal risk sharing mechanisms \cite{BHATTAMISHRA2010923,DELPIERRE2016282}.

As society develops, the modern financial system has introduced index insurance products, an innovative risk management tool. Index insurance provides economic compensation to the insured through predefined triggering conditions (such as specific meteorological indicators or the frequency of natural disasters) \cite{Jensen2017,Miranda2012,santos2021dynamics}. Its advantages include a fast, transparent claims process, and the characteristic of not relying on individual loss assessments. Through standardized risk assessment, index insurance not only reduces moral hazards but also ensures the rational distribution of insurance funds \cite{Chantarat2007,Miranda2011}. Although index insurance theoretically offers an efficient mechanism for risk transfer, it still faces several challenges in practical application. Basis risk is one of the main problems. This risk refers to the possibility that when the insured actually suffers a loss, they may not receive compensation from index insurance due to the non-fulfillment of specific trigger conditions, even if their loss is quite evident. This could lead to insufficient economic protection for the insured, thereby failing to meet their needs for risk coverage \cite{BUDHATHOKI20191,CLEMENT2018845,CONRADT2015106,Elabed2013,LICHTENBERG2022102883,RIGO2022102237}.

Informal risk sharing mechanisms and formal index insurance products have become the main options for farmers to cope with natural disasters. Previous studies usually considered the limited evolutionary dynamics of individual decisions under a single risk management strategy, although empirical studies have pointed out that farmers tend to rely more on informal mechanisms under high basis risk \cite{Mobarak2013}. However, it remained unclear how the risk environment affects individual behavioral decisions when the two mechanisms are working simultaneously. In addition, for insurance companies, how to reasonably price index insurance to promote sales and sustainability is still an important issue to be solved urgently \cite{DARON201476,TAIB201222}. Evolutionary game theory provides a new perspective and method for studying this complex problem \cite{Christopher2023Similarity,Badal2021auto,liu2019evolutionary,liu2023coevolutionary,Liu2024Fixation,ozkan2016application,PERC20171,PERC2010109,santos2016evolution,smith1982evolution,szolnoki2014cyclic,tori2022study,wang2024evolutionary,wang2024paradigm,HUA2024121579}.

The aim of this work is to explore the evolutionary process of individuals' decisions among choosing an informal risk sharing strategy, purchasing index insurance products, and not purchasing any insurance. To this end, we construct an evolutionary game model to gain an in-depth understanding of how factors such as basis risk, risk sharing ratio, and index insurance pricing collectively influence individuals' preferences for different insurance strategies. Moreover, this work proposes an optimized index insurance pricing scheme intended to maintain the attractiveness of the insurance products to individuals and maximize the profits of insurance companies, thereby ensuring the long-term stability and sustainability of the insurance market.
    
\section{Model and Methods}
We consider a population of $Z$ individuals who are all exposed to the threat of natural disasters, with a probability $p$ of occurrence. When an individual suffers from a disaster, a proportion $\alpha$ of their total wealth $w$ will be lost. Individuals can adopt two insurance strategies for risk transfer: the first is to participate in the construction of an informal risk sharing pool to reduce expected losses by collectively sharing the risks among participants (denoted as $S$), and the second is to purchase index insurance products offered by insurance companies (denoted as $I$). Additionally, individuals can choose not to participate in any insurance plan, bearing the loss caused by disasters on their own (denoted as $A$).

Individuals participating in the informal risk sharing pool are required to contribute a fraction, denoted as $\delta_{1}$, of their total wealth to establish a collective fund aimed at mitigating the losses associated with disasters. In the event of a disaster, the accumulated fund is distributed equally among the affected members. Conversely, if no disaster occurs, the complete amount of the fund is returned to the contributing members. It is noteworthy that such risk sharing pools operate on a small scale, typically at the level of villages or small towns, which facilitates the process of loss assessment.

Individuals opting to purchase index insurance are required to pay an insurance premium, denoted as $c$. The insurance company will make payments based on a predetermined index, compensating the individual's total loss once the conditions for payment are triggered. Let $q$ represent the probability of payment being triggered. We define $r$ as the probability of a disaster occurring without triggering an index insurance payment. Consequently, we can derive the probability of the following events: $p-r$ represents the probability of a disaster occurring alongside a triggered payment; $q+r-p$ reflects the likelihood of no disaster occurring while still triggering an index insurance payment; and $1-q-r$ signifies the probability of neither a disaster occurring nor an index insurance payment being triggered \cite{santos2021dynamics}.

To mitigate the damage caused by basis risk, index insurance incorporates an additional risk sharing mechanism. Individuals who purchase the insurance must also contribute an additional proportion, $\delta_{2}$, of their wealth to establish a separate risk sharing pool. If a catastrophe occurs and the index insurance does not provide compensation, the funds allocated to this pool will be distributed equally to all insured members affected by the catastrophe who have not received compensation. Conversely, if no claims are made, the pooled funds are returned in full to all contributing investors.

We assume that the insurance plan is offered to a group of $N$ individuals. Suppose there are $k$ individuals choosing to establish an informal risk sharing pool, $l$ individuals purchasing index insurance, and the remaining $N-k-l$ individuals not participating in any insurance activities. In this context, the payoffs for individuals establishing the informal risk sharing pool, denoted as $\pi_{S}(k)$, for those purchasing index insurance, denoted as $\pi_{I}(l)$, and for those not participating in any insurance, denoted as $\pi_{A}$, are defined as follows:
\begin{align}
	&\pi_{S}(k)  =\frac{1}{k} \sum_{h=0}^{k}\binom{k}{h} p^{h}(1-p)^{k-h} Q_S(h), \label{eq:pi_S}\\
	&\pi_{I}(l)  =\frac{1}{l} \sum_{u+v+m+n=l} P_I(u, v, m, n) Q_I(u, v, m, n), \label{eq:pi_I}\\
	&\pi_{A}  =p U((1-\alpha) w)+(1-p) U(w), \label{eq:pi_A}
\end{align}
where the utility function $U(\cdot)$ follows a constant relative risk aversion (CRRA) form, represented as  $U(x)=\frac{x^{1-\gamma}}{1-\gamma}$ \cite{huang2008}. Here, the parameter $\gamma$ indicates the degree of risk aversion of individuals; a higher value of $\gamma$ signifies a stronger aversion to risk. The function $Q_S(h)$ quantifies the total utility for participants in the informal risk sharing pool, with $h \in \{0,1,\dots,k\}$ indicating the count of pool members, out of $k$, who suffer a disaster. In contrast, for the $l$ individuals opting for index insurance, the joint probability distribution $P_I(u,v,m,n)$, along with the total utility function $Q_I(u,v,m,n)$, is established based on the distribution of the four possible outcomes ($u+v+m+n=l$) associated with index insurance: $u$ denotes those afflicted by disaster and receiving compensation, $v$ represents individuals spared from disaster without any payout, $m$ encapsulates those erroneously compensated despite no disaster, and $n$ includes those who endure a disaster but fail to receive the corresponding payout. These functions $Q_S(h)$, $P_I(u, v, m, n)$, and $Q_I(u, v, m, n)$ are defined as follows:
\begin{align}
	&Q_S(h) = 
	\begin{cases}
            h U\left((1-\alpha) w - \delta_{1} w + \frac{k \delta_{1} w}{h}\right) \\
            \quad + (k - h) U\left(w - \delta_{1} w\right), & \text{if } h \neq 0, \\
            k U(w), & \text{if } h = 0,
        \end{cases} \\ 
	 &P_I(u, v, m, n) \notag\\ 
     &= \binom{l}{u} \binom{l-u}{v} \binom{l-u-v}{m}  \notag\\
     & \quad \times (p - r)^{u} (1 - q - r)^{v} (q + r - p)^{m} r^{n}, \\
	&Q_I(u, v, m, n)  \notag\\
    &= 
	\begin{cases}
		(u + v) U\left(w - c - \delta_{2} w\right) \\ \quad+ m U\left(w - c + \alpha w - \delta_{2} w\right) \\
		\quad + n U\left((1 - \alpha) w - c - \delta_{2} w + \frac{l \delta_{2} w}{n}\right), & \text{if } n \neq 0, \\
		(u + v) U(w - c) + m U(w - c + \alpha w), & \text{if } n = 0.
	\end{cases}\label{q(u,v,m,n)}
\end{align}

We adopt the Markov process to simulate the dynamic changes of system state in a finite population \cite{Kim2024finite}. We define the state space  $\mathcal{S}=\left\{\mathbf{s_{1}}, \mathbf{s_{2}}, \ldots, \mathbf{s_{\frac{(Z+1)(Z+2)}{2}}}\right\}$, where each state  $\mathbf{s_{i}}$ corresponds to a specific strategy configuration tuple $\left(i_{S}, i_{I}\right)$ \cite{liu2024evolution}. In this configuration, $i_{S}$ and $i_{I}$ denote the number of individuals choosing the informal risk sharing pool strategy $(S)$ and the index insurance strategy $(I)$, respectively, and  $Z-i_{S}-i_{I}$ represents the number of individuals not opting any insurance strategy. Under a given configuration tuple $\mathbf{s_i}=\left(i_S, i_I\right)$, and combining Eq. \eqref{eq:pi_S} - \eqref{q(u,v,m,n)}, individual fitness based on these three different strategies is denoted as $f_{S}(i_S, i_I)$, $f_{I}(i_S, i_I)$ and $f_{A}(i_S, i_I)$, calculated as follows:
\begin{align}
	f_{S}(i_S, i_I) & =\sum_{k=0}^{N-1} \sum_{l=0}^{N-1-k}  \frac{\binom{i_S-1}{k}\binom{i_I}{l}\binom{Z-i_S-i_I}{N-1-k-l}}{\binom{Z-1}{N-1}}\pi_{S}(k+1), \\
	f_{I}(i_S, i_I) & =\sum_{k=0}^{N-1} \sum_{l=0}^{N-1-k}  \frac{\binom{i_S}{k}\binom{i_I-1}{l}\binom{Z-i_S-i_I}{N-1-k-l}}{\binom{Z-1}{N-1}}\pi_{I}(l+1), \\
	f_{A}(i_S, i_I) & =\pi_{A}.
\end{align}

To investigate the dynamic decision-making processes underlying individuals' strategic choices in natural disaster risk management, we adopted a model based on individual fitness to simulate the behavior of individuals adjusting their strategies. The probability of an individual switching from strategy $X$ to strategy $Y$ is estimated using the Fermi function \cite{szabo1998evolutionary}, represented by the following formula:
\begin{eqnarray}
	P_{\left(X \leftarrow Y\right)} & = & \frac{1}{1+e^{\beta\left(f_{X}-f_{Y}\right)}},
\end{eqnarray}
where $f_{X}$ and $f_{Y}$ denote the fitness associated with adopting strategies $X$ and $Y$, respectively. The parameter $\beta$ describes the influence of fitness differences on the probability of strategy switching. Given the stochastic nature of individual behavior, we introduce a probability of behavior mutation $\mu$ to account for uncertainty and non-rational decision factors. Consequently, under the configuration tuple $\mathbf{s_i}$, the one-step transition probability for an individual moving from strategy $X$ to $Y$ is given by
\begin{align}\label{trans_XY}
	T_{X \rightarrow Y}^{\mathbf{s_i}}  =  (1-\mu)\left[\frac{i_{X}}{Z} \frac{i_{Y}}{Z-1} \frac{1}{1+e^{\beta\left(f_{X}-f_{Y}\right)}}\right]+\mu \frac{i_{X}}{\left(d-1\right) Z},
\end{align}
where $i_{X}$ and $i_{Y}$ represent the number of individuals in the current configuration tuple $s_i$ that have adopted the strategies $X$ and $Y$, respectively, and the variable $d$ represents the number of strategies that an individual can choose from.

Next, we present the transition probabilities for individuals moving from the current state $\left(i_{S}, i_{I}\right)$ to other possible states as follows:
\begin{eqnarray*}
	\begin{array}{ll}
		{\left(i_{S}, i_{I}\right) \rightarrow\left(i_{S}+1, i_{I}\right)}=T_{A \rightarrow S}^{\mathbf{s_i}}, \\ {\left(i_{S}, i_{I}\right) \rightarrow\left(i_{S}-1, i_{I}\right)}=T_{S \rightarrow A}^{\mathbf{s_i}}, \\
		{\left(i_{S}, i_{I}\right) \rightarrow\left(i_{S}, i_{I}+1\right)}=T_{A \rightarrow I}^{\mathbf{s_i}}, \\ {\left(i_{S}, i_{I}\right) \rightarrow\left(i_{S}, i_{I}-1\right)}=T_{I \rightarrow A}^{\mathbf{s_i}}, \\
		{\left(i_{S}, i_{I}\right) \rightarrow\left(i_{S}+1, i_{I}-1\right)}=T_{I \rightarrow S}^{\mathbf{s_i}}, \\ {\left(i_{S}, i_{I}\right) \rightarrow\left(i_{S}-1, i_{I}+1\right)}=T_{S \rightarrow I}^{\mathbf{s_i}}, \\
		{\left(i_{S}, i_{I}\right) \rightarrow\left(i_{S}, i_{I}\right)}=1-\underset{U \neq U^{\prime}}{\sum} T_{U \rightarrow U^{\prime}}^{\mathbf{s_i}}, &
	\end{array}
\end{eqnarray*}
where $U$, $U^{\prime} \in\{S, I, A\}$. In order to quantify the configuration that is most likely to reach the end of the next time step after the system leaves the current configuration, based on equation \eqref{trans_XY}, we defined the selection gradient $\vec{\nabla}({\mathbf{s_i}})$ as follows:
\begin{eqnarray*}
	\vec{\nabla}({\mathbf{s_i}}) & = & \left[\begin{array}{l}
		T_{S \rightarrow I}^{\mathbf{s_i}}+T_{A \rightarrow I}^{\mathbf{s_i}}-T_{I \rightarrow S}^{\mathbf{s_i}}-T_{I \rightarrow A}^{\mathbf{s_i}} \\
		T_{I \rightarrow S}^{\mathbf{s_i}}+T_{A \rightarrow S}^{\mathbf{s_i}}-T_{S \rightarrow I}^{\mathbf{s_i}}-T_{S \rightarrow A}^{\mathbf{s_i}}
	\end{array}\right] .
\end{eqnarray*}

Finally, we calculate the average adoption frequency of the three strategies to quantify the prevalence of each strategy in the population:
$$\bar{p}_{S}=\sum_{\mathbf{s} \in \mathcal{S}} \frac{\mathbf{s}_{S} \bar{p}_{\mathbf{s}}}{Z}, \quad \bar{p}_{I}=\sum_{\mathbf{s} \in \mathcal{S}} \frac{\mathbf{s}_{I} \bar{p}_{\mathbf{s}}}{Z}, \quad \bar{p}_{A}=\sum_{\mathbf{s} \in \mathcal{S}} \frac{\mathbf{s}_{A} \bar{p}_{\mathbf{s}}}{Z},$$
where  $\mathbf{s}_{S}$, $\mathbf{s}_{I}$, and $\mathbf{s}_{A}$ denote the number of individuals choosing the informal risk sharing pool, purchasing index insurance, and not participating in any insurance under state $\mathbf{s}$, respectively. $\bar{p}_{\mathbf{s}}$ represents the probability of state $\mathbf{s}$ in the stationary distribution.



\section{Results}

\begin{figure*}[tbhp]
	\centerline{\includegraphics[width=1\textwidth]{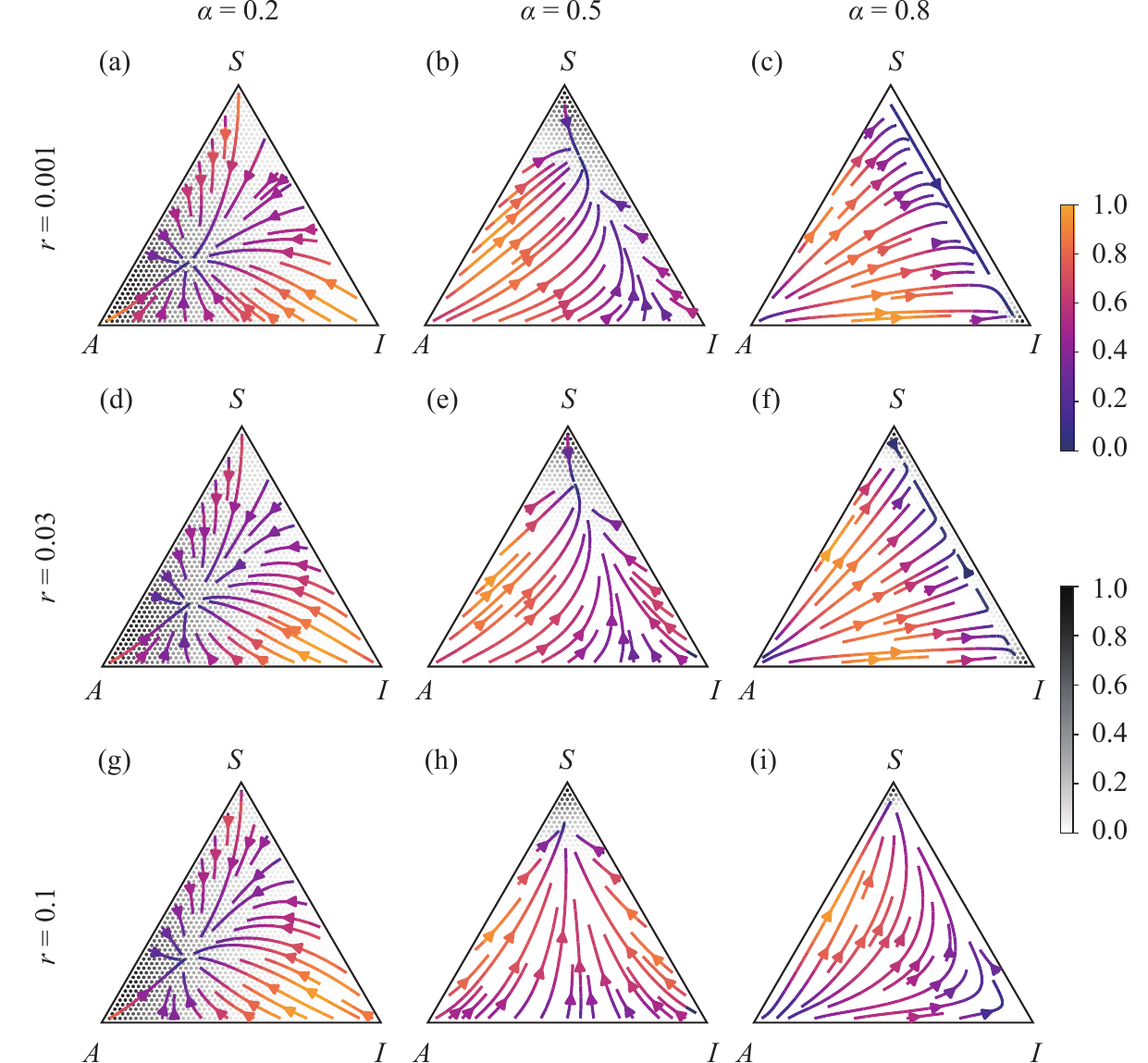}}
	\caption{\textbf{The evolutionary dynamics of three risk management strategies $(S, I, A)$ under different risk loss ratios $\alpha$ and basis risk $r$.} Panels (a-i) show the system's evolutionary paths of the system under different levels of basis risk and risk loss ratios, where the gray scale bar on the right indicates the relative probability of the system reaching different steady states, with darker shades representing higher staying probabilities. The color bar on the right indicates the relative strength of the selection gradient, and the color changes intuitively reflect the variation in selection gradients under different states. Each row has the same level of basis risk, and each column has the same risk loss ratio. Remaining parameters for panels (a-i): $w=1$, $p=0.2$, $q=0.2$, $\alpha=0.8$, $\gamma=0.8$, $\delta_{1}=0.1$, $\delta_{2}=0.05$, $\beta=10, c=\alpha w q+0.01$, $Z=50$, $N=40$, and $\mu=0.02$.}
	\label{FIG: alpha}
\end{figure*}

\begin{figure*}[tbhp]
	\centerline{\includegraphics[width=1\textwidth]{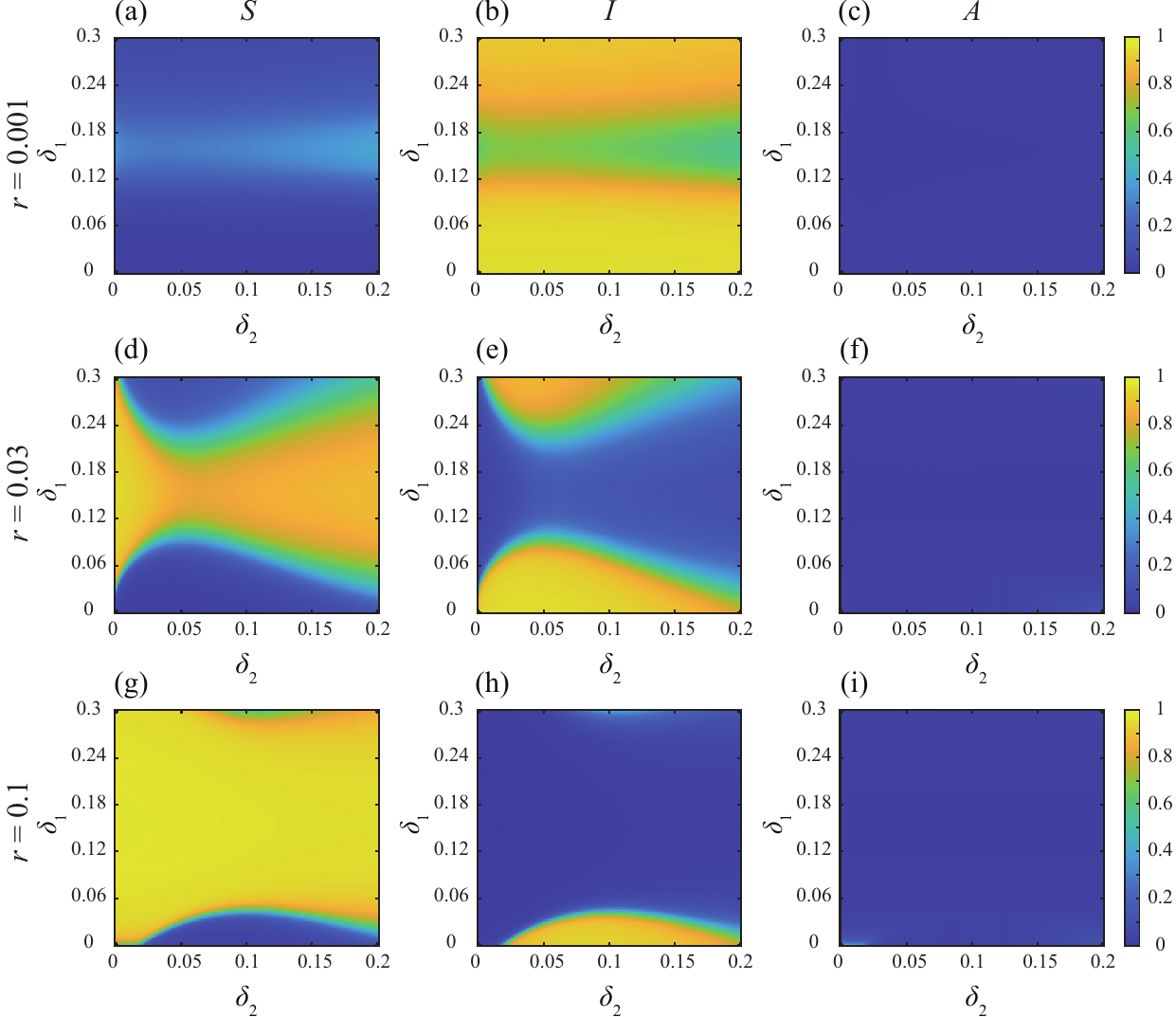}}
	\caption{\textbf{The average adoption rates of three risk management strategies $(S, I, A)$ under different risk sharing ratios $\delta_{1}$ and $\delta_{2}$.} Panels (a-i) uses different colors to show the relative sizes of the average adoption rates of the three strategies under different $\delta_{1}$ and $\delta_{2}$ parameter conditions. The color bar on the far right represents an increase in the average adoption rate from bottom to top. Remaining parameters for panel (a-i): $w=1$, $p=0.2$, $q=0.2$, $\alpha=0.8$, $\gamma=0.8$, $\beta=10$, $c=0.17$, $Z=50$, $N=40$, and $\mu=0.02$.}
	\label{FIG: delta}
\end{figure*}

We first examined the impact of the basis risk $r$ and the risk loss ratio $\alpha$ on individual decision-making behavior (see \cref{FIG: alpha}). In an environment where the loss ratio is low ($\alpha=0.2$), even if the basis risk of the index insurance is low, individuals usually have the ability to bear losses independently and therefore tend not to participate in any form of insurance plan (see \cref{FIG: alpha}(a), (d), and (g)). However, when the loss ratio rises to a moderate level ($\alpha=0.5$), informal risk sharing mechanisms begin to operate. Regardless of the level of basis risk, individuals can effectively offset losses from unfortunate events by joining an informal risk sharing pool, showing a clear preference for informal risk sharing mechanisms (see \cref{FIG: alpha}(b), (e), and (h)). As the loss ratio increases further ($\alpha=0.8$), the design advantages of index insurance products gradually become apparent, and it is at this point that basis risk has a significant impact on individual choice preferences. In particular, when the basis risk is low ($r=0.001$), index insurance tends to become the preferred choice for individuals seeking to reduce potential risks, due to the higher credibility of the insurance payout mechanism (see \cref{FIG: alpha}(c)). However, as the basis risk increases ($r=0.03$), the credibility of the insurance payouts decreases, reducing the attractiveness of index insurance products. In this scenario, informal risk sharing strategies have significant advantages. Compared to not participating in an insurance plan, establishing an informal risk sharing pool allows individuals to collectively share the risk of natural catastrophes internally, thereby significantly increasing their resilience (see \cref{FIG: alpha}(f) and (i)). Notably, the system can exhibit bistable outcomes. Depending on the different initial strategy choices, the system may eventually evolve into different states ($S$ or $I$)(see \cref{FIG: alpha}(f)). In environments with high basis risk ($r=0.1$), this informal risk sharing mechanism serves as an effective risk management tool, attracting more individuals to adopt it (see \cref{FIG: alpha}(i)).

\begin{figure*}[tbhp]
	\centerline{\includegraphics[width=1\textwidth]{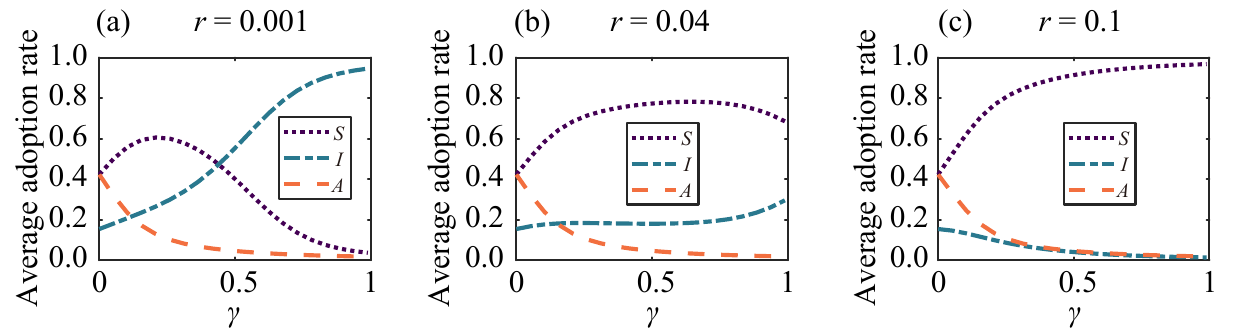}}
	\caption{\textbf{The average adoption rates of three risk management strategies $(S, I, A)$ under different degrees of risk aversion $\gamma$.} Remaining parameters for panel (a-c): $w=1$, $p=0.2$, $q=0.2$, $\alpha=0.8$, $\delta_{1}=0.1$, $\delta_{2}=0.05$, $\beta=10$, $c=0.17$, $Z=50$, $N=40$, and $\mu=0.02$.}
	\label{FIG: gamma}
\end{figure*}

In a further study, we examine the impact of the informal risk sharing ratio $\delta_{1}$ and the index risk sharing ratio $\delta_{2}$ on the take-up rates of strategies $S$, $I$, and $A$ (see \cref{FIG: delta}). The findings further highlight the centrality of basis risk in the promotion of index insurance, and also reveal the importance of setting a reasonable risk sharing ratio, which has a significant impact on an individual's choice of insurance strategy. Specifically, when the basis risk is small ($r=0.001$), index insurance products show significant advantages and become a popular risk management tool. However, the risk diversification effect of introducing informal risk sharing pools and setting appropriate informal risk sharing ratios may reduce the attractiveness of index insurance products (see \cref{FIG: delta}(a)-(c)). When the basis risk is moderate ($r=0.03$), a reasonable informal risk sharing ratio can significantly enhance the effectiveness of informal risk sharing pools, making them the preferred strategy for individuals to deal with natural disaster risks. However, too high or too low an informal risk sharing ratio can reduce the attractiveness of informal risk sharing pools and thus create a market opportunity for index insurance. In this context, appropriate index risk sharing ratios can enhance the competitiveness of index insurance and make them a more attractive option (see \cref{FIG: delta}(d)-(f)). As basis risk rises further ($r=0.1$), even with the presence of an index risk sharing mechanism, the attractiveness of index insurance products is significantly diminished, making it difficult to compete effectively with informal risk sharing strategies (see \cref{FIG: delta}(g)-(i)). It is worth noting that the average adoption rate of non-participation in any insurance strategy is consistently low, regardless of changes in basis risk. Even in scenarios where index insurance is less attractive, informal risk sharing pools, with their effective risk diversification mechanisms, can still attract individuals and play a key role.

We subsequently investigate the impact of risk aversion $\gamma$ under different levels of basis risk $r$ on the adoption rates of strategies $S$, $I$, and $A$. Our results show that the level of risk aversion of individuals has a significant differential effect on their choice of behavior strategies at different levels of basis risk (see \cref{FIG: gamma}). In scenarios with lower basis risk ($r=0.001$), the index insurance becomes more attractive to individuals as risk aversion increases due to its excellent compensation ability (see \cref{FIG: gamma}(a)). However, as basis risk increases, the likelihood of a mismatch between payouts and actual losses (mismatched compensation) increases for index insurance, leading to a decrease in its attractiveness (see \cref{FIG: gamma}(b) and (c)). Under these circumstances, the informal risk sharing pool strategy begins to show its advantages, especially for those with higher risk aversion. By diversifying risk, the risk sharing pool becomes a more reliable choice.

In the insurance market, the premium pricing strategy is crucial to the profit model of insurance companies. On one hand, if premiums are set too low, insurance companies may fail to cover their costs and thus face the risk of losses. On the other hand, excessively high premiums may inhibit the purchasing intentions of potential consumers, thereby reducing the market penetration rate of insurance products. Therefore, finding a premium level that not only attracts a sufficient number of individuals but also ensures the profitability of insurance companies requires careful consideration. 

\begin{figure*}[tbhp]
	\centerline{\includegraphics[width=1\textwidth]{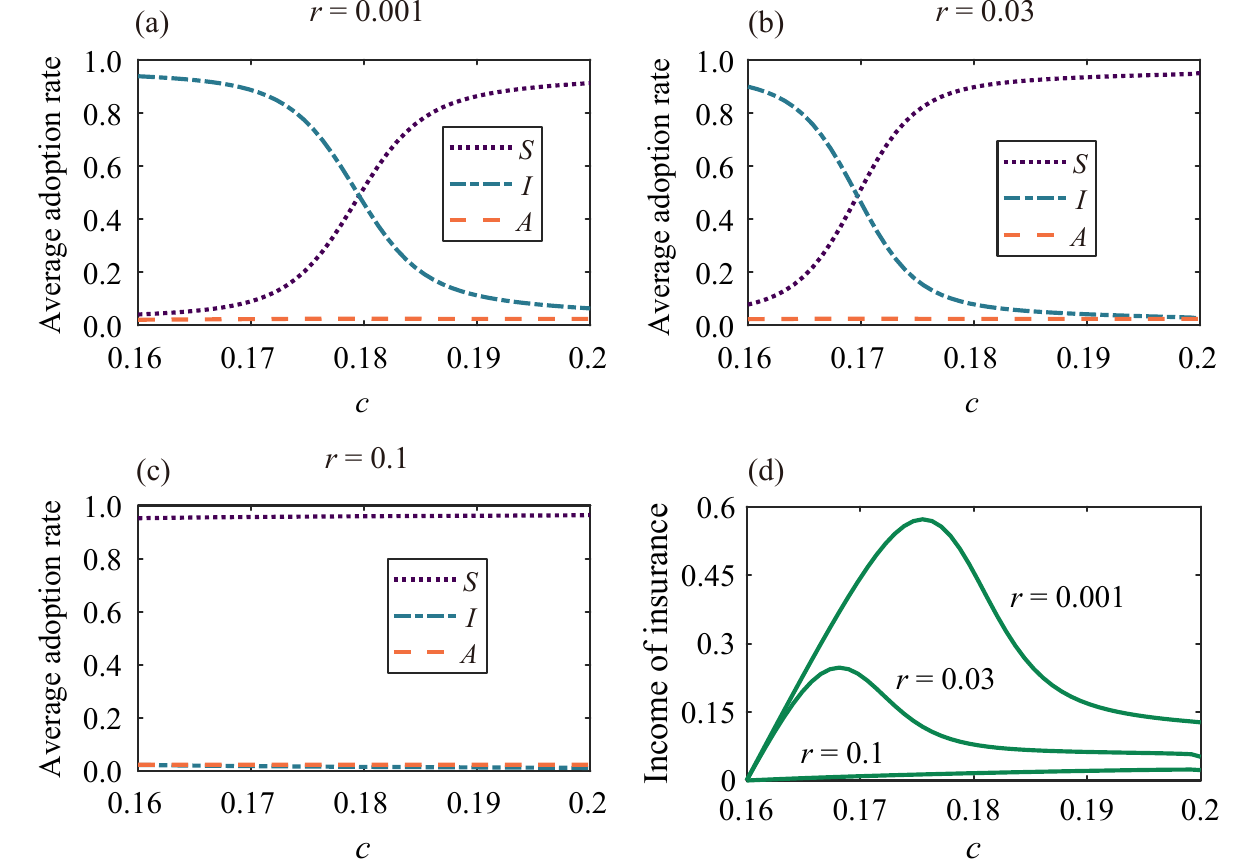}}
	\caption{\textbf{The average adoption rates of three strategies $(S, I, A)$ and the expected profit of insurance company under different insurance premiums $c$.} Panels (a)-(c) demonstrate the impact of premiums on individual strategy choices under different basis risks. Panel (d) shows the potential profit of the insurance company at different premiums under different basis risks. The remaining parameters for panels (a)-(d) are $w=1$, $p=0.2$, $q=0.2$, $\alpha=0.8$, $\gamma=0.8$, $\delta_{1}=0.1$, $\delta_{2}=0.05$, $\beta=10$, $Z=50$, $N=40$, and $\mu=0.02$.}
	\label{FIG: c}
\end{figure*}

To quantitatively analyze the relationship between premium pricing and the expected profits of insurance companies, we define the expected profit $\left(\bar{\pi}_{C}\right)$ of the insurance company as $\bar{\pi}_{C}  =  \bar{p}_{I} Z(c-\alpha w q)$. In \cref{FIG: c}, we illustrate the impact of the premium $c$ on the average adoption rate of these three strategies $(S, I, A)$, and based on this, we calculate the expected profit of insurance companies at different premium levels. We find that when premiums are priced too low, insurance companies may achieve a higher sales volume, but due to the lower profit per policy, the overall profit remains unsatisfactory. Conversely, when premiums are set too high, although the profit per policy increases, a significant decrease in sales volume leads to a reduction in the overall profit of the insurance company (see \cref{FIG: c}(a)-(c)). In \cref{FIG: c}(d), we show the expected profits of insurers under different premium pricing. The results of the study show that there exists an optimal premium pricing that maximizes the insurer's profit in all three basis risk environments. At this level of pricing, insurers are able to achieve an optimal balance between attracting consumers and maintaining profitability. However, with the gradual increase in basis risk, the insurer's expected maximum profit shows a downward trend. This result highlights that accurate index setting to minimize basis risk is crucial for insurers, not only to enhance product attractiveness, but also to improve profitability stability and thus long-term sustainability.

\begin{figure*}[tbhp]
        \centerline{\includegraphics[width=1\textwidth]{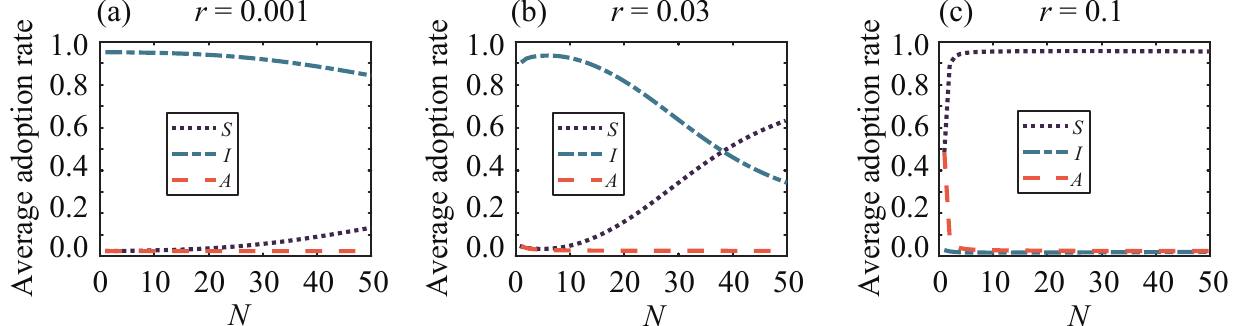}}
	\caption{\textbf{The average adoption rates of three strategies $(S, I, A)$ under different group size $N$.} Panels (a)-(c) show the effect of group size $N$ on individual decisions under different basis risks $r$. The remaining parameters for panels (a)-(c) are $w=1$, $p=0.2$, $q=0.2$, $\alpha=0.8$, $\gamma=0.8$, $\delta_{1}=0.1$, $\delta_{2}=0.05$, $\beta=10$, $Z=50$, $c=0.17$, and $\mu=0.02$.}
	\label{FIG: N}
\end{figure*}

Subsequently, we examine the effect of group size $N$ on individual strategy choice (see \cref{FIG: N}). We find that in all three basis risk environments, a sustained increase in group size leads to a gradual decrease in the attractiveness of index insurance and a significant increase in the adoption of the informal risk sharing pool strategy. This phenomenon can be attributed to the fact that larger group sizes allow for the construction of a more stable and well-funded informal risk sharing pool. In this case, individuals tend to share the risk they face by joining the pool, which significantly increases the attractiveness of the strategy.

Importantly, we also investigate the evolutionary dynamics of a system with only two strategies, $S$ and $A$ (Supplementary Information, Section 1), demonstrating the effectiveness of the risk sharing pool under appropriate risk environments (see Fig. S1). Specifically, it shows that having a significant advantage under high disaster occurrence rates and high disaster loss rates, compared to not participating in any insurance plan. In the face of potential losses, participants often prefer to establish an internal structure to achieve higher expected profits through collective risk sharing among individuals. The study further examines the impact of risk aversion on the participation rate in the risk sharing pool (see Fig. S2); as individual risk aversion increases, the attractiveness of the risk sharing pool strengthens, prompting more individuals to adopt this strategy.

Finally, we further investigate the evolutionary dynamics of the system when only strategies $S$ and $I$ are present (Supplementary Information, Section 2), revealing that basis risk and the risk loss ratio are two key factors affecting the market performance of index insurance products (see Fig. S3 and Fig. S4). When basis risk is low, index insurance is more popular due to its superior compensation performance; however, as basis risk increases, its market attractiveness declines. Additionally, an increase in risk aversion leads individuals to prefer avoiding insurance products with higher uncertainty. Under conditions of high basis risk and low risk loss ratios, as risk aversion increases, individuals are more likely to opt for an informal risk sharing pool strategy. Conversely, under conditions of high risk loss ratios and low basis risk, an increase in risk aversion results in a higher adoption rate of index insurance.

\section{Conclusions}

In the field of disaster risk management, index insurance strategies and informal risk sharing pool strategies are two common mechanisms for risk sharing. These strategies reduce the losses individuals may suffer from natural disasters \cite{santos2021dynamics}. Index insurance relies on insurance companies to trigger payouts based on a predetermined index \cite{pacheco2016evolutionary}, while informal risk sharing pools are based on the mutual bearing and sharing of losses among group members. When these two strategies coexist, the insurance selection behavior of individuals becomes complex but highly valuable for research, as they directly affect the market acceptance and sustainability of index insurance products. In this work, we have conducted a comprehensive analysis of key factors such as basis risk, risk sharing ratios, risk aversion, and loss ratios, examining how these variables influence individuals' strategy selection decisions when facing natural disaster risks. The findings reveal the relative advantages of index insurance and informal risk sharing pool strategies, emphasizing the applicability and effectiveness of various strategies in risk management.

We have shown that basis risk and the risk loss ratio are significant factors affecting the sales of index insurance products. When basis risk is low and the risk loss ratio is high, index insurance is widely welcomed due to its strong reputation and third-party reimbursement capability. In this scenario, individuals are more inclined to utilize index insurance against potential risk losses. However, when the risk loss ratio is at a moderate level, the potential of risk sharing pools begins to emerge. In this context, individuals can better reduce their personal risk by joining a collective risk sharing mechanism, making risk sharing pools a more popular choice. When the risk loss ratio is low, individuals may lose only a part of their initial endowment even if they encounter disasters. At this time, they tend to choose not to participate in any insurance plan, bearing the loss alone. This behavior indicates that the demand for insurance largely depends on the severity of the loss. Therefore, index insurance is particularly suitable for situations with severe disaster losses, such as the impact of natural disasters like floods and droughts on agricultural production. In these cases, index insurance provides an effective risk management tool to help farmers mitigate potential losses when facing significant risks when the basis risk is low. Thus, it is crucial to design and promote index insurance products tailored to this specific market to address the challenges and risks brought about by severe disasters.

Moreover, properly adjusting the risk sharing ratio is critical for enhancing the attractiveness of insurance strategies. A moderate risk sharing ratio can not only optimize the risk diversification effect but also significantly increase the strategy's appeal to individuals. This adjustment helps make insurance products more aligned with participants' needs and expectations while ensuring effective risk sharing. At the same time, accurately assessing individuals' risk aversion is equally crucial for insurance companies in designing and selling index insurance products. An individual's risk aversion directly affects their acceptance and willingness to participate in specific insurance products. Therefore, understanding consumers' risk aversion can provide insurance institutions with a more targeted basis for product design, thereby increasing market share.

Finally, the study suggests that insurance companies can find an effective balance between attracting consumers and maintaining profitability by optimizing premium settings. This finding provides important guidance for market strategies, emphasizing the significance of precise pricing in insurance product design. Reasonable premium pricing can enhance the product's market competitiveness and increase consumers' purchase intentions while contributing to the company's long-term sustainable profitability. Therefore, precise premium setting should become a crucial part of insurance companies' strategic planning to achieve the dual goals of economic benefits and consumer satisfaction.

Although this study investigates the potential irrational behaviors of individuals by incorporating behavioral mutations and offers insights into the individual decision-making process, real-world behaviors are influenced by various irrational factors such as emotions and social relationships \cite{lerner2015emotion,martin2011influence}. Moreover, discrepancies exist between the assumed homogeneity of individuals in the study and the significant differences observed in reality \cite{lieberman2005evolutionary,ni2023heterogeneous,pinheiro2012local}. Therefore, future research focusing on the dynamics of individual strategy evolution in heterogeneous complex networks will enhance the accuracy and applicability of the models, enabling a more comprehensive explanation and prediction of individual behaviors.

\end{multicols}

\enlargethispage{20pt}

\ethics{The study involved no human or animal subjects, thus exempt from ethical approval requirements.}

\dataccess{All code is placed on the Zenodo platform (https://zenodo.org/records/15229757).}

\aucontribute{\textbf{Lichen Wang:} Conceptualization, Methodology, Software, Writing – original draft, Writing – review \& editing. \textbf{Yuyuan Liu, Shijia Hua, Zhengyuan Lu and Liang Zhang:} Methodology, Software, Writing – review \& editing. \textbf{Linjie Liu and  Attila Szolnoki:} Supervision, Methodology, Validation.}
\disclaimer{We have not used AI-assisted technologies in creating this article.}
\competing{We declare we have no competing interests.}

\funding{This work was funded by the Humanities and Social Sciences Research Planning Fund of the Ministry of Education (No. 24XJC630006), the National Natural Science Foundation of China (Nos. 62306243, and 62406255), and China Postdoctoral Science Foundation (Certificate Number: 2024M762633). A. S. was supported by the National Research, Development and Innovation Office (Grant No. K142948).}


\printbibliography



\end{document}